\documentclass{vietnam}

\usepackage[style=numeric,sorting=none]{biblatex}
\renewbibmacro{in:}{}
\def\be{\begin{equation}}
\def\ee{\end{equation}}
\def\bea{\begin{eqnarray}}
\def\eea{\end{eqnarray}}

\usepackage[nolist]{acronym}
\begin{acronym}

\acro{GW}{Gravitational wave}
\acro{BBH}{Binary Black Holes}
\acro{AdV}{Advanced Virgo}
\acro{MC}{Mode-cleaner}
\acro{aLIGO}{Advanced LIGO}
\acro{GR}{General Relativity}
\acro{IM}{Input Mirror}
\acro{EM}{End Mirror}
\acro{ITF}{interferometer}
\acro{SRM}{Signal Recycling Mirror}
\acro{PRM}{Power Recycling Mirror}
\acro{BS}{Beam-Splitter}
\acro{BNS}{Binary Neutron Star}
\acro{SNR}{Signal-to-Noise-Ratio}
\acro{QN}{Quantum Noise}
\acro{SN}{Shot Noise}
\acro{RPN}{Radiation Pressure Noise}
\acro{FIS}{Frequency-Independent Squeezing}
\acro{FDS}{Frequency-Dependent Squeezing}
\acro{TM}{Test Masses}
\acro{FC}{Filter Cavity}
\acro{AdV+}{Advanced Virgo+}

\end{acronym}  

\usepackage[binary-units]{siunitx}
\DeclareSIUnit[number-unit-product = {}]{\inchQ}{''}

\graphicspath{{Figures/}}

\usepackage{indentfirst}

\newcommand{\Photo}{\includegraphics[height=35mm]{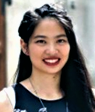}}

\begin{document}
\vspace*{4cm}
\title{STATUS OF THE ADVANCED VIRGO GRAVITATIONAL-WAVE DETECTOR  }

\author{ CATHERINE NGUYEN on behalf of the Virgo Collaboration }

\address{Universit\'e de Paris, CNRS, AstroParticule et Cosmologie, F-75013 Paris, France}

\maketitle\abstracts{ On September 2015, a century after Einstein's predictions of their existence, the first \acp{GW} direct detection was performed by LIGO. On August 17, 2017,  the two Advanced LIGO and the Advanced Virgo interferometers detected a \ac{GW} produced by two merging neutron stars. The subsequent localization of the source in the sky, thanks to the presence of a third detector, led to the detection of the electromagnetic counterpart and follow-up of the event by roughly 70 electromagnetic and neutrino telescopes. After the first two data taking runs (O1 and O2), the LIGO-Virgo network detected 11 \acp{GW} from 10 \ac{BBH} and one \ac{BNS}. On April 1, 2019, Advanced Virgo and Advanced LIGO started their third observing period (O3). After an introduction on \ac{GW} detection, I will give an overview on the Advanced Virgo detector design, with a description of the technical choices made before O3 and their consequences on the detector sensitivity. Finally, I will describe the planned upgrades for the \ac{AdV+} project.
}

\section{Introduction}

\par In the last 50 years, \ac{GW} detectors based on modified Michelson interferometers with kilometric arm length have been developed, and are now forming an international network composed of the two \ac{aLIGO} detectors\cite{2015} (LIGO Hanford, in the state of Washington and LIGO Livingston in the state of Louisiana) and the \ac{AdV} detector (near Pisa, in Italy)\cite{Acernese_2014}. Moreover, the KAGRA\cite{Akutsu_2020} detector (at the Kamioka mine in Japan)
had nearly reached the \ac{BNS} range  \footnote{The \ac{BNS} range is a standard figure of merit for the sensitivity of the interferometer, which is the averaged distance at which it can detect a \ac{BNS} merger composed of two 1.4 ${M}_{\odot}$ neutron stars, with a signal-to-noise ratio of 8 for sources uniformally distributed over the sky with random inclination and polarization angles.} threshold of 1 Mpc\cite{kagra} to join the recent science run (O3) but did not have time before its suspension due to the COVID-19 crisis. A smaller detector, GEO 600, is also operational in Germany\cite{Dooley_2015}.
\par On September 14, 2015, during the first LIGO observational run (O1), the two LIGO detectors achieved the first GW direct detection \cite{PhysRevLett.116.061102}, measuring \ac{GR} in the unexplored regime of strong field and relativistic speed. This observation demonstrates the existence of stellar mass black holes more massive than $25  M_{\odot}$, and the possibility for binary systems of these objects to merge within the Hubble time. 

\par On August 14, 2017, during the second science run O2, the \ac{aLIGO}-\ac{AdV} network made their first triple detection of a BBH merger (GW170814)\cite{PhysRevLett.119.141101} and three days later, it observed \acp{GW} emitted by a \ac{BNS} merger (GW170717) in coincidence with a $\gamma$-ray burst (GRB 170817A) detection by the Fermi Gamma Ray Burst Monitor and INTEGRAL\cite{PhysRevLett.119.161101},
confirming the hypothesis that short-$\gamma$-ray bursts are linked to \ac{BNS} mergers. Moreover, the delay on the arrival times was used to test \ac{GR} and put strong constraints on the \ac{GW} speed \cite{Abbott_2017}. 
In addition to the detection of GRB 170817A, the LIGO-Virgo localization of the source in the sky allowed a follow-up by roughly 70 electromagnetic and neutrino observatories, identifying the host galaxy and a \textit{kilonova}\cite{multimessenger}.
This unprecedented joint \ac{GW} and electromagnetic observation gave new insights into astrophysics, dense matter under extreme conditions and tests of \ac{GR}. Moreover, the joint measurement of redshift by the host galaxy identification and by the \ac{GW} luminosity distance enabled a novel measurement of the Hubble constant\cite{PhysRevLett.119.161101}. 

The two LIGO-Virgo observational runs (O1 and O2) resulted in the detection of \acp{GW} from 10 \ac{BBH} mergers and from 1 \ac{BNS} merger, reported in the first catalog of gravitational-wave transient sources GWTC-1\cite{PhysRevX.9.031040}. 
After detector upgrades of \ac{AdV} and \ac{aLIGO}, the third observational run (O3) started on April, 1, 2019 with an average BNS range of 110 Mpc and 130 Mpc for \ac{aLIGO} Hanford and \ac{aLIGO} Livingston respectively. 
\ac{AdV} started with a range of about 45 Mpc and recently reached 60 Mpc. The O3 run was suspended on March 27, 2020, a month before the official closing time, due to sanitary reasons.

\begin{figure}[ht]
\begin{minipage}[t]{0.55\linewidth}
\centering
\includegraphics[width=\columnwidth]{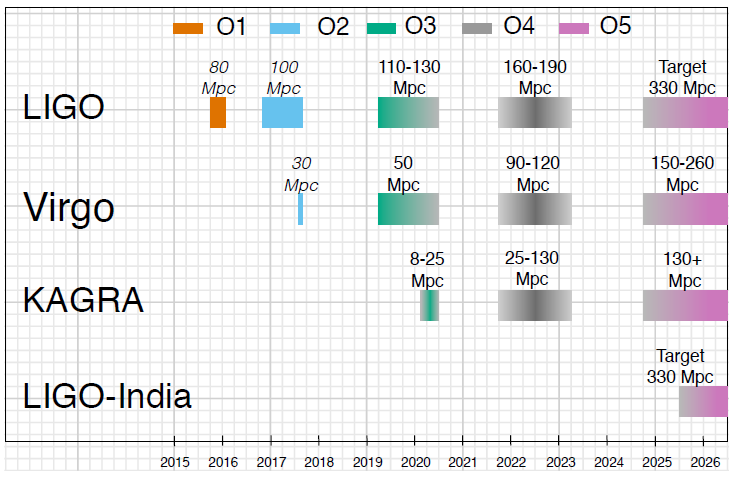}
   \caption{Observing scenarios with targeted sensitivities for aLIGO, AdV, KAGRA and LIGO-India GW detection network for the coming years\protect\cite{collaboration2013prospects}.}
\label{fig:OR}
\end{minipage}
\hfill
\begin{minipage}[t]{0.40\linewidth}
\includegraphics[width=\columnwidth]{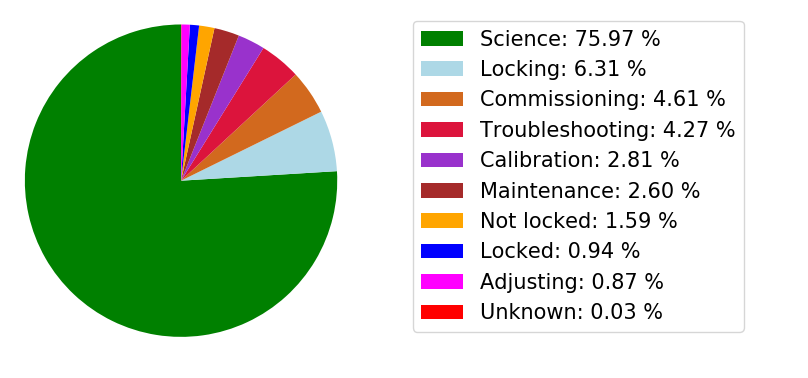}
    \caption{Status of \ac{AdV} for the O3a and O3b, without accounting for the commissioning break (credit: LIGO-Virgo collaboration).}
\label{fig:dutyAdV}
\end{minipage}
\end{figure}

In the next years, \ac{AdV} and \ac{aLIGO} will see upgrades alternated with observational periods. Figure~\ref{fig:OR} shows current and planned sensitivity for past and future observation runs.

In Section 2, the current design of \ac{AdV} is presented, Section 3 describes the changes made between O2 and O3, Section 4 details the upgrade to a squeezed vacuum states, Section 5 gives a brief summary of the O3 results; and finally, Section 6 presents the \ac{AdV+} project, the planned upgrades for \ac{AdV}.

\section{Advanced Virgo design}

 The initial Virgo detector was operational from 2007 until 2011 when it was upgraded to \ac{AdV}\cite{acernese2019increasing}. The goal was to increase the sensitivity by an order of magnitude, which would allow an increase in detection rate of three orders of magnitude.    

\par
\begin{figure}[ht]
\begin{minipage}[t]{0.50\linewidth}
\includegraphics[width=\columnwidth]{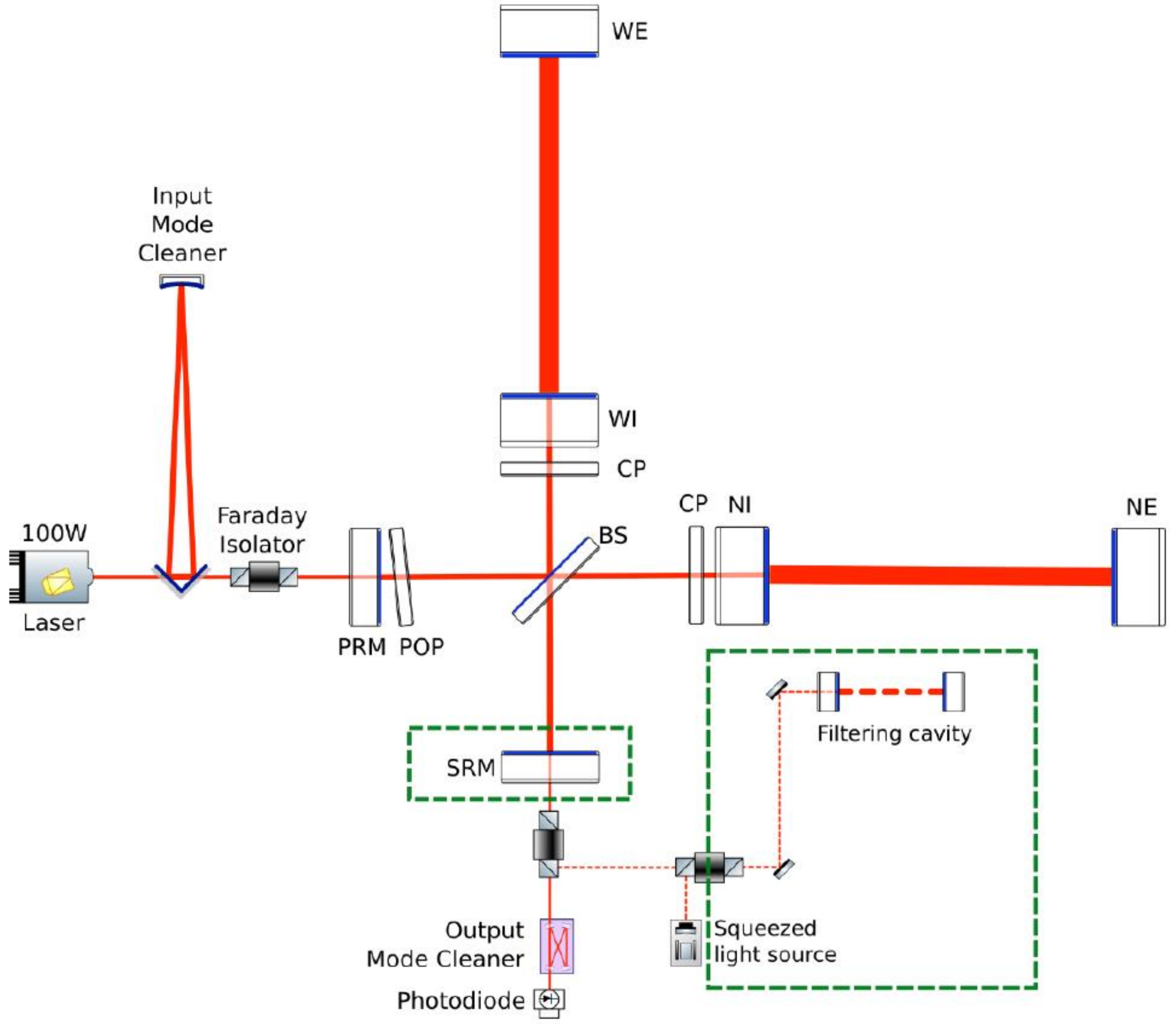}
    \caption{ Simplified optical layout of the \ac{AdV} detector in its current configuration and with some planned upgrades for the Advanced Virgo+ (AdV+) project\protect\cite{AdV+}. Among these upgrades, the installation of the filter cavity and the Signal-Recycling Mirror (SRC) are framed in dotted green.} 
\label{fig:ITF}
\end{minipage}
\hspace{0.3cm}
\begin{minipage}[t]{0.45\linewidth}
\centering
\includegraphics[width=\columnwidth]{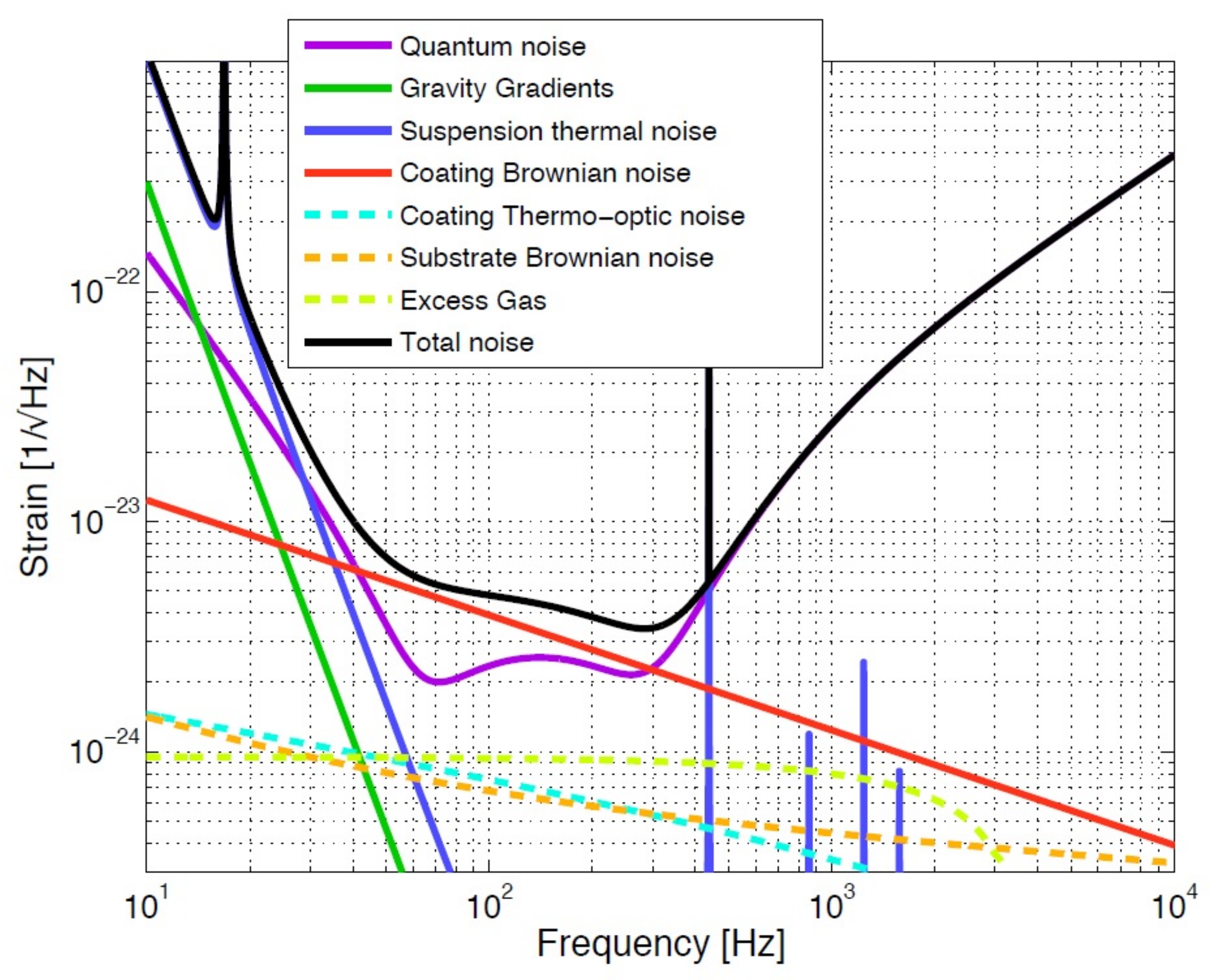}
   \caption{Design sensitivity curve of \ac{AdV} with its noise budget\protect\cite{Acernese_2014}. In the low frequency region (up to \SI{40}{\hertz}), the sensitivity is dominated by Newtonian noise, seismic noise and suspension thermal noise. The mid-frequency range is mainly limited by coating thermal noise and the high-frequency region (above \SI{300}{\hertz}) is dominated by shot noise.
   }
\label{fig:sensitivity}
\end{minipage}
\end{figure}

 The detector is a power-recycled Michelson, whose simplified optical layout is depicted in Figure~\ref{fig:ITF}, with two Fabry-Perot cavities in each arm. The resonant Fabry-Perot cavities have a finesse of about 450 and enable an increase of the effective optical length of roughly 300. Moreover, a power recycling cavity about \SI{12}{\meter} long, formed by a \ac{PRM} placed between the laser and the \ac{BS}, enhances the circulating optical power by roughly 36. 
\\The \ac{ITF} is illuminated by a Nd:YAG laser source at \SI{1064}{\nano\meter} which delivers an input power of \SI{26}{\watt} stabilized in frequency, power and position thanks to active filtering and a 144-meter suspended \ac{MC} cavity with a finesse of 1000. The \ac{GW} signal is detected at the anti-symmetric port by DC detection,  after filtering by two output \ac{MC} cavities  to remove both the spatial high-order modes and the RF frequencies that control the auxiliary degrees of freedom of the detector.

Figure~\ref{fig:sensitivity} shows the planned sensitivity of \ac{AdV} with the main noise sources. In the low frequency region (up to \SI{40}{\hertz}), the sensitivity is dominated by Newtonian noise, seismic noise and suspension noise. The mid-frequency range is mainly limited by the suspension thermal noise (the suspension thermal noise and the coating Brownian noise) and the high-frequency region (above \SI{300}{\hertz}) is dominated by shot noise (whose mitigation is described in Section \ref{sec:squeezing}), the high-frequency component of the quantum noise. 

\subsection{Seismic noise and seismic isolation}

In order to filter the seismic noise (due to terrestrial seismic motion, wind and human activity), the \ac{ITF} \ac{TM} (4 mirrors of the Fabry-Perot cavities), \ac{PRM} and the \ac{BS} are all suspended with a multi-pendulum chain, acting as horizontal filters above their resonance frequency (about 0.5 Hz). The chain is suspended using a pre-isolator: an inverted pendulum that guarantees isolation below 1 Hz and damping of the resonances of the chain \cite{Acernese_2014}.
The residual seismic noise, after filtering by this “super-attenuator" ssytem, is shown in the dark blue curve of Figure~\ref{fig:sensitivity}. 

In order to attenuate the consequences of the seismic noise (for example, scattered light), all the sensitive optical components, for example the optical benches with the input-output auxiliary optics (i.e. input and ouput \acp{MC}, mode-matching telescopes) are suspended using special suspension systems, similar in concept to the super-attenuator\cite{Acernese_2014}.

\subsection{Mirrors, monolithic suspensions and thermal noise}
The thermal noise in \ac{AdV} has two main sources: the suspension wires and the mirror coating.
The mirrors thermal noise (coating and substrate) can be divided into thermo-elastic effects (thermal fluctuations coupling with a non-zero thermal expansion coefficient of the material), Brownian motion (kinetic energy of the mirrors atoms at temperature $T$), and thermo-refractive fluctuations (thermodynamical fluctuations coupled with variation of material refractive index with temperature)\cite{Acernese_2014,acernese:hal-01965299}.
The dominant source is from Brownian thermal noise of the coatings. 
To mitigate this Brownian thermal noise, the multi-layer coatings are made with fused silica (low index material) and Ti doped $\text{Ta}_{\text{2}}\text{O}_{\text{5}}$ (high index material) which has the lowest mechanical losses\cite{Granata:19}.
\par The “monolithic suspension" is composed of \SI{400}{\micro\meter}-diameter $\text{SiO}_{\text{2}}$ fibers. Fused silica, a high strength material with low mechanical losses whose intrinsic dissipation is about three orders of magnitude lower than steel\cite{acernese:hal-01965299}, is tailored for the suspension of these heavy mirrors (42 kg) in order to reduce the effect of the suspension thermal noise. 

\section{Improvement of AdV between O2 and O3}

Between O2 and O3, several improvements were made to increase the \ac{AdV} sensitivity: the injection of a squeezed vacuum reaching 2-3 dB as described in Section \ref{sec:squeezing}; the replacement of the steel \ac{TM} suspension wires with SiO4 fibers, lowering the dissipation in the pendulum and hence the suspension thermal noise; the installation of a \SI{100}{\watt} laser amplifier and increase of the laser input power from \SI{10}{\watt} to \SI{26}{\watt}; the installation of additional baffles in several locations to mitigate against scattered light; and the refinement of global alignment control at higher bandwidth than in O2\cite{collaboration2020gw190425}.
All these improvements enabled an increase in \ac{BNS} range from almost 30 Mpc (end of O2) to 60 Mpc (reached on the beginning of February 2020).  

\section{Implementation of squeezing vacuum in AdV} \label{sec:squeezing}

\ac{QN} originates in the quantum nature of light: it is a consequence of the Heisenberg Uncertainty Principle\cite{caves1981quantum}.
\ac{QN} is composed of \ac{SN}, dominating above \SI{100}{\hertz}, and \ac{RPN}, dominating below $\sim$ 100 Hz. As the \ac{SN} equivalent strain sensitivity is inversely proportional to laser input power, the natural solution to reduce this source of noise is to increase the laser power. This has been done before and during O3, and allowed an increase sensitivity at high frequency. 
\\ However, the power increase causes thermal effects in the mirror and consequently thermal aberrations due to power absorption.
\par In 1981, Caves proposed an alternative technique called \ac{FIS} to decrease shot noise without increasing laser power. Since quantum noise is generated by vacuum fluctuations entering from the unused port of the \ac{BS}, he\cite{caves1981quantum} proposed a replacement of the standard vacuum by squeezed vacuum states that have lower fluctuations in the quadrature aligned with the \ac{GW} signal. These squeezed states will however increase the \ac{RPN}, but in the current detectors this is barely visible as the low-frequency part of the sensitivity is dominated by technical noise sources.

\par
\begin{figure}[ht]
\begin{minipage}[t]{0.40\linewidth}
\centering
\includegraphics[width=\columnwidth]{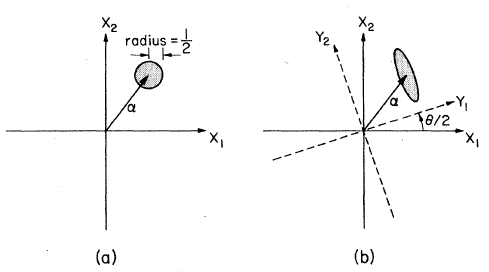}
    \caption{Representation in a phase-space quadratures plane\protect\cite{caves1981quantum} of (a) a coherent state of light represented by an error circle, and (b) a squeezed state of light represented by an error ellipse.} 
\label{fig:SqzEllipse}
\end{minipage}
\hspace{0.3cm}
\begin{minipage}[t]{0.55\linewidth}
   \centering
   \includegraphics[width=\columnwidth]{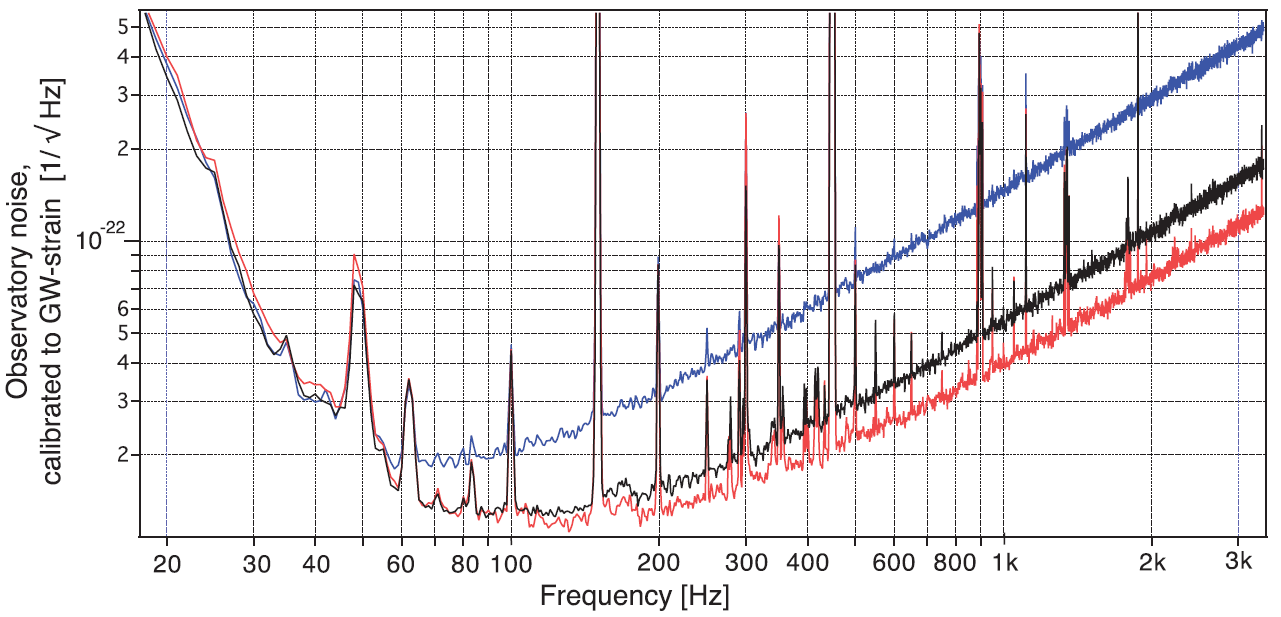}
   \caption{\ac{AdV} sensitivity curves in different conditions of squeezed light injection\protect\cite{acernese2019increasing}. The black trace corresponds to the measured sensitivity without squeezing, the red and blue traces correspond to the measured sensitivity with squeezing and anti-squeezing respectively.}
\label{fig:SqzSensitivity}
\end{minipage}
\end{figure}

 Figure~\ref{fig:SqzEllipse} represents the uncertainty area in the phasor (phase-amplitude) plane with a circle for a coherent state and an ellipse for a squeezed state. 
Squeezed states of light and vacuum are produced using nonlinear optical processes and generated in an external in-air optical bench, built by the GEO600 group in Hannover and installed in Virgo in January 2018. The \ac{FIS} technique is routinely used by \ac{AdV}~\cite{acernese2019increasing} and \ac{aLIGO}\cite{PhysRevLett.123.231107} with a quantum enhancement between \SI{100}{\hertz} and \SI{3.2}{\kilo\hertz} of up to 3.2 $\pm$ 0.1 dB and around 3 dB, respectively. It is equivalent to an increase of the input power by a factor 2. Figure~\ref{fig:SqzSensitivity} shows the strain sensitivity of the \ac{AdV} detector without squeezing (black curve) and with squeezing (red curve), an improvement in sensitivity from 52 Mpc to 55 Mpc\cite{acernese2019increasing}. 

\section{Main results on AdV performance during O3}

\begin{figure}[ht]
\centering
\includegraphics[width=0.55\columnwidth]{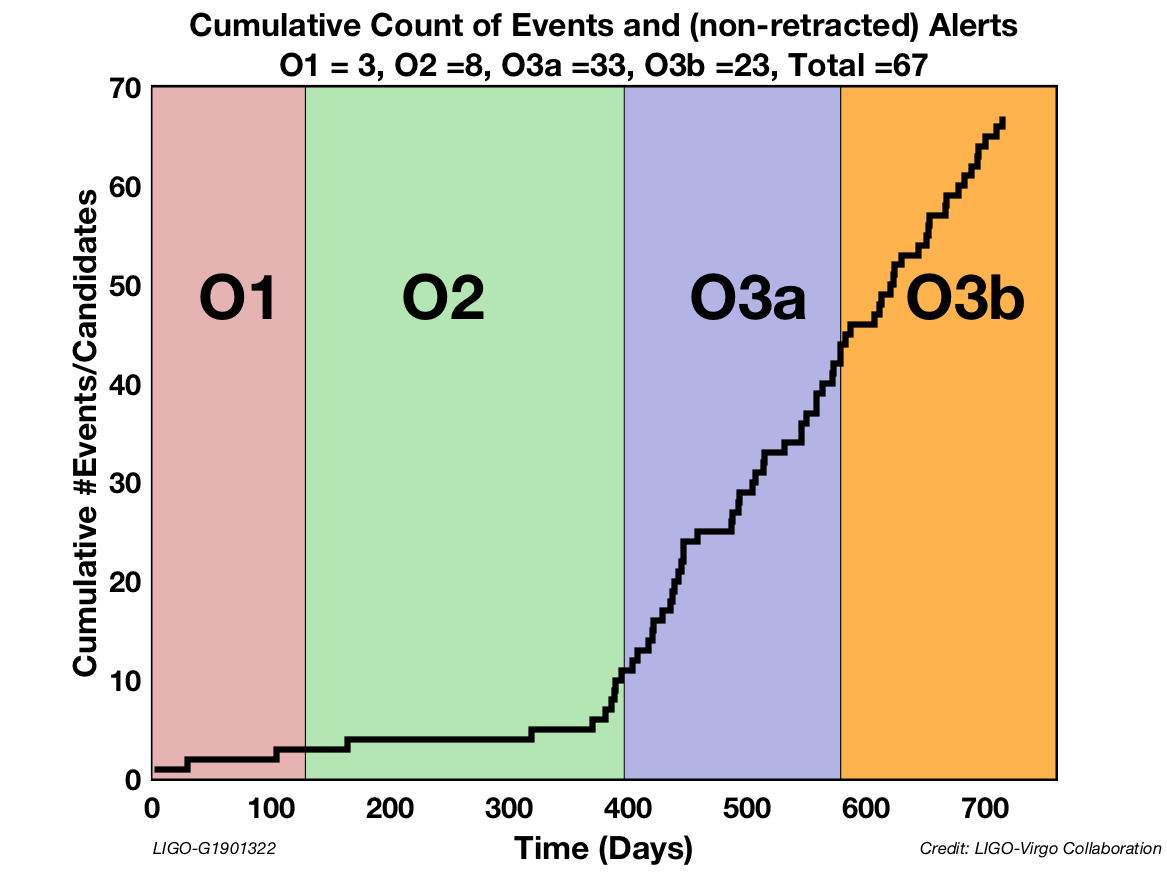}
   \caption{Cumulative events (O1 and O2) or candidates (O3) versus time for the three observing runs (credit: LIGO-Virgo collaboration).}
\label{fig:cumevents}
\end{figure}

\par The O3 run is divided into 2 periods called O3a and O3b. This two periods are separated by a commissioning break during October 2019. As a result, before O3 suspension, \ac{aLIGO} detectors have an average BNS range of about 120 Mpc and 135 Mpc for Hanford\cite{LH} and Livingston\cite{LL} respectively. For \ac{AdV}, the BNS range reached 60 Mpc \cite{virgo}, which is more than two times the BNS range at the end of O2. The chart in Figure~\ref{fig:dutyAdV} shows the different status of \ac{AdV} during O3, illustrating that the duty cycle was nearly 76\% for the cumulative 
period.
\\For the whole O3 run before its suspension (from April 1, 2019 to March 27, 2019), the network duty operational state was as follows: the 3 detectors were online together 47.4\% of the time, 2 detectors were online 36\% of the time, only one detector was online 13.3\% of the time and no \ac{ITF} was locked during 3.3\% of the time.
\par From Figure~\ref{fig:cumevents}, it can be deduced that roughly one GW candidate per week was identified with 56 total candidates by April 1, 2020. The increase in detection rate demonstrates that the upgrades made between O2 and O3 were clearly fruitful. A significant \ac{GW} signal was detected on April 25, 2019, originating most likely from a \ac{BNS} merger with total mass significantly larger than any other known \ac{BNS} system\cite{collaboration2020gw190425}.
The LIGO-Virgo GW alerts are public and posted on a dedicated website \cite{graceDB}.

\section{Plan for Advanced Virgo+}
The next upgrade of \ac{AdV} is called \ac{AdV+} and is divided in two phases.
Phase 1 will be implemented between O3 and O4 and will mainly concern changes related to quantum noise reduction. Phase 2, scheduled between O4 and O5, will mainly focus on thermal noise reduction. The target sensitivity for the improvements of phase 1 is a BNS range of the order of 100 Mpc for coalescing \ac{BNS} and above 200 Mpc for phase 2\cite{AdV+}.

\subsection{Phase 1}

\begin{figure}[ht]
\centering
\includegraphics[width=0.8\columnwidth]{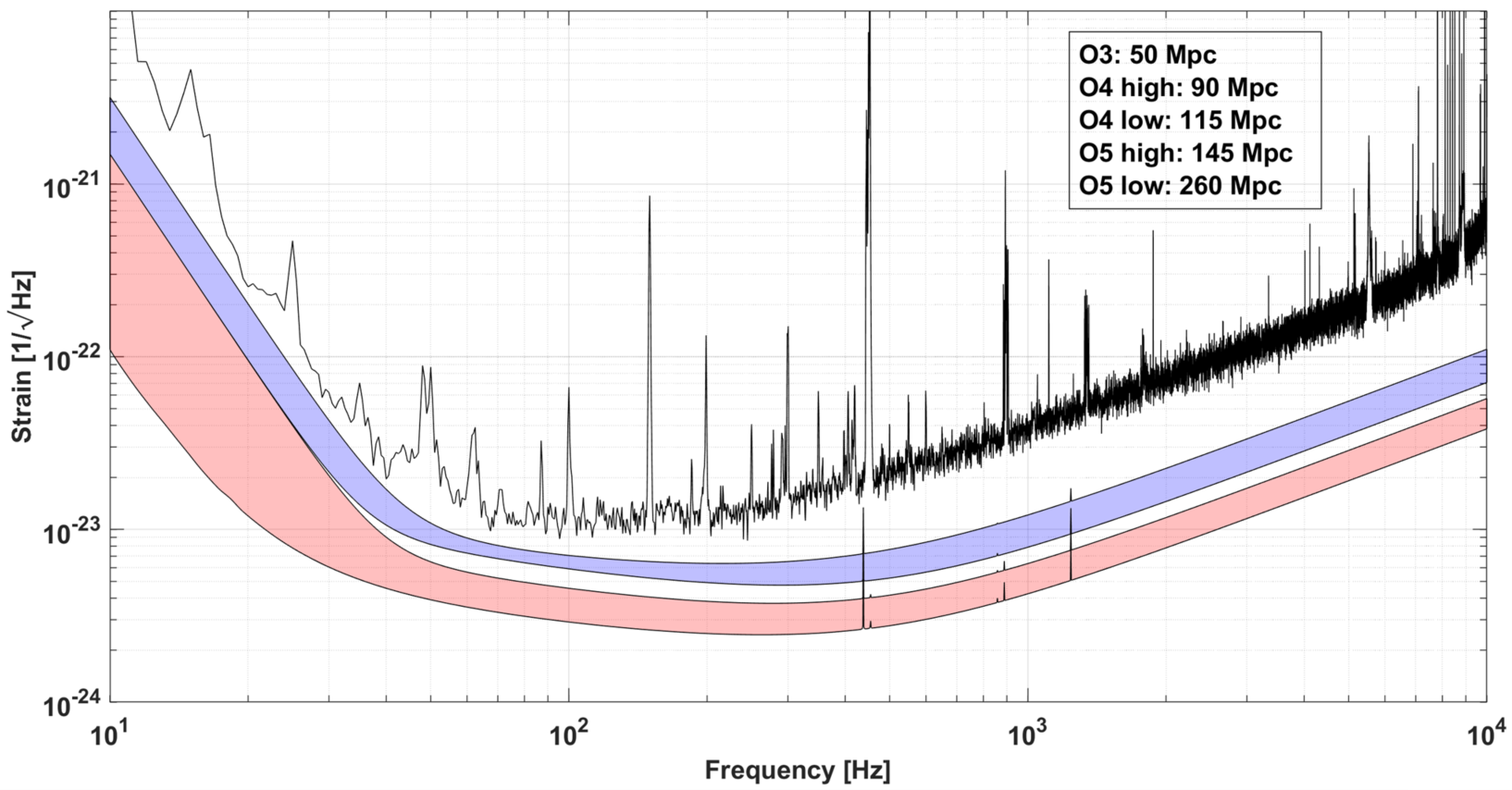}
   \caption{AdV+ anticipated sensitivities for O4 and O5\protect\cite{Flaminio_2019}. As a comparison, the sensitivity at the beginning of O3 is shown in gray.}
\label{fig:SqzAnticipatedSensitivity}
\end{figure}

The main changes proposed for phase 1 involve a signal recycling mirror installation, laser input power increase (from the current \SI{26}{\watt} to \SI{50}{\watt}), \ac{FDS} with a filter cavity and the Newtonian noise cancellation technique. During this phase, the mirrors currently installed as well and the beam geometry in the \ac{ITF} arms, will be kept, so that the mirror thermal noise will remain at the current level. 
\par Signal recycling is a method to optimize the response of the detectors to a GW by enlarging their bandwidth \cite{AdV+}. 
The \ac{SRM}, depicted in Figure~\ref{fig:ITF} (framed in dotted-green), will be placed between the \ac{BS} and the interferometer output, improving the sensitivity at high frequencies. This signal recycling cavity will enable the change of the shape of the sensitivity curve by tuning its length and finesse, optimizing for different astrophysical sources. 
The whole configuration is called \textit{dual-recycled Fabry-Perot Michelson interferometer}\cite{acernese:hal-01965299}.
\\ Furthermore, a complex signal is needed to acquire the locking of the interferometer, involving auxiliary lasers to lock the 3-km cavities\cite{AdV+}.

\par As shown in Figure~\ref{fig:SqzAnticipatedSensitivity}, the technical noises will be reduced, leading the \ac{RPN} to dominate the sensitivity of \ac{AdV+} at lower frequencies, and hence the \ac{FIS} technique will be inadequate (it increased the \ac{RPN}, see Section \ref{sec:squeezing}).
To address this problem, the squeezing ellipse  should be rotated in a frequency-dependent way.
As a consequence, the \ac{FDS} technique with the use of a very long Fabry-Perot cavity of \SI{285}{\meter}, referred to as \textit{filter cavity} and planned for \ac{AdV+}. As shown in Figure~\ref{fig:ITF}, the vacuum produced by the squeezed light source will be injected into the filter cavity where it experiences a differential phase change depending on the cavity parameters, which leads to the rotation of the squeezing ellipse in the phase-amplitude quadrature plane\cite{AdV+} and therefore a broadband reduction of the quantum noise. An experiment using a suspended 300 m filter cavity, similar to the one planned for KAGRA, \ac{AdV} and \ac{aLIGO}, sucessfully demonstrated this \ac{FDS} technique\cite{zhao2020frequencydependent}.  

\par The Newtonian noise in the detector will become a limiting noise between about \SI{10}{\hertz} and \SI{20}{\hertz}. This frequency range is interesting from a scientific point of view, for example to detect intermediate-mass black-hole binaries. Hence the testing of Newtonian noise cancellation techniques with the deployment of an array of seismometers, which will be installed at each end building and the central building where the dominant contributor to the seismic noise is located\cite{AdV+,Tringali_2019}.

\subsection{Phase 2}

Phase 2 focuses on thermal noise reduction with an increase of mirror weight beam size. A larger payload and increase of the laser power to 200W is also planned. R\&D is also currently underway by a collaboration of several laboratories to address losses by mirror coatings. 

\par To cope with the increased impact of radiation pressure fluctuations on \acp{TM}, the beam on the \acp{EM} will be enlarged to a radius of \SI{10}{\centi \meter}  so the increased heat from the higher cavity is spread over a large area and heavier mirrors (100 kg) reduce mirror motion due to the radiation pressure.
The two \ac{EM} will have a diameter of \SI{55}{\centi \meter}, representing an increase of \SI{60}{\percent}. The super-attenuator load capability will be increased by, for example, adjusting the handling tool for larger mirrors and upgrades on seismic filter to cope with larger payload weight. 

\section{Summary}

\ac{AdV} contributed to the the O2 LIGO-Virgo data taking. In addition to the observation of \acp{GW} from a total of seven \acp{BBH} merger, the first \ac{BNS} detection (\textit{GW170817}) on August 17, 2017, with the observation of the electromagnetic counterpart associated to this event, marked the beginning of the multi-messenger astronomy. 
Between O2 and O3, an increase of the \ac{AdV} sensitivity of around a factor two has been obtained thanks to several upgrades, including the use of squeezed states of vacuum and the replacement of steel suspensions with fused silica fibers. As a result, the \ac{ITF} remained stable with a duty cycle higher than 75\% for the whole O3 run until its suspension in March 2020 due to the global pandemic.
\par At the time of the writing, 56 \ac{GW} candidates have been identified in O3. All alerts during O3 were public, with the first published analysis by the collaboration on the particular event of \textit{GW190425}.
Thanks to the huge effort of commissioning, noise hunting and detector characterisation, the current BNS range finished at 60 Mpc for \ac{AdV}. The \ac{AdV+} project plans to further increase the \ac{AdV} science reach, with a planned \ac{BNS} range above 200 Mpc.

\printbibliography
\end{document}